# The thermally activated motion of vortex bundles and the anomalous Hall effect in type-II conventional and high-Tc superconductors

**Wei Yeu Chen*, Ming Ju Chou**

Department of Physics, Tamkang University, Tamsui 25137, Taiwan

**Abstract**

The anomalous Hall effect in type-II conventional and high-Tc superconductors is investigated based on the proposed novel theory of the thermally activated motion of vortex bundles over a directional-dependent energy barrier. Our calculations demonstrate clearly that the anomalous Hall effect is induced by the competition between the Magnus force and the random collective pinning force of the vortex bundle. The Hall as well as the longitudinal resistivity for constant temperature and constant applied magnetic field of type-II superconducting films and bulk materials are calculated. The reentry phenomenon is also investigated. All the results are in good agreement with the experiments.

* Corresponding author. Tel.: 886-2-26206871; fax: 886-2-26206871.
E-mail: wychen@mail.tku.edu.tw (W.Y. Chen)

## 1. Introduction

The vortex dynamics in type-II superconductors, discovered by Abrikosov [1], has been focused tremendously theoretic and experimental efforts [2-25], especially the most appealing and confusing phenomenon-the anomalous Hall effect (AHE)- in type-II conventional and high-Tc superconductors[14-25], since the Hall resistivity was first measured by Niessen and Staas in 1965 [2], later the Hall anomaly was observed by Van Beelen et al. [3]. So far, there is no generally satisfactory explanation for this amazing anomaly, the origin of this Hall anomaly still remains unsolved. In this paper, we develop a novel theory for the thermally activated motion of vortex bundle over the directional-dependent energy barrier formed by the potential generated by the Magnus force and the collective pinning force, and the potential barrier generated by a myriad of randomly distributed strong pinning sites inside the bundle [4, 6-9]. Applying this theory, we successfully explain the remarkable Hall anomaly. In accordance with our theory, the AHE is indeed induced by the competition between the Magnus force and the random collective pinning force of the vortex bundle.

The rest of this paper is organized as follows. In Section 2, the theory of thermally activated motion of vortex-bundle is developed; the Hall and longitudinal resistivities are calculated. In Sections 3 and 4, the AHE for type-II superconducting bulk materials and thin films are investigated respectively. In Section 5, the reentry phenomenon for the AHE is discussed. In Sections 6 and 7, a general discussion and concluding remarks are conveyed.

## 2. The thermally activated motion of vortex bundles and the Hall and longitudinal resistivities

Realizing the fact that quenched disorder destroys the long-range order of the flux lattice (FLL), after which only short-range order, the vortex bundles, prevails [5-6]. In this section we develop a theory for thermally activated motion of vortex bundles over the directional-dependent energy barrier formed by the Magnus force, the random collective pinning force, and the potential barrier generated by the randomly distributed strong pinning sites inside the bundle. The energy barrier is directional-dependent due to the presence of Magnus force and random collective pinning force, namely, the energy barrier is different when the direction of activated motion is different. Our results demonstrate that the AHE is induced by the competition between Magnus force and random collective pinning force of the vortex-bundle.

*2.1. Calculation of the coherent frequency of the vortex bundle by random walk theorem*

It is well understood that [10] the vortex line oscillates inside the potential barrier due to thermal agitations. The thermal oscillation frequency of the individual vortex inside the potential barrier, by identifying the oscillating energy of the vortex line with the thermal energy, can easily be obtained as,

$$\nu = \bar{\nu}\sqrt{T} \quad , \tag{2.1}$$

with $\bar{\nu} = \frac{1}{\pi A}\sqrt{\frac{k_B}{2m}}$, where $A$ is the average amplitude of the oscillation, $k_B$ is the Boltzmann constant and $m$ is the mass of the vortex line [11]. The oscillations of each individual vortex inside the vortex bundle are not coherent, namely, their oscillations are at random. In order to obtain the coherent oscillation frequency $\nu_C$ of the vortex bundle as a whole, the frequency $\nu$ in Eq. (2.1) must be divided by the square root of the number of vortices in the bundle $\sqrt{N}$ ,

$$\nu_C = \frac{\nu}{\sqrt{N}} = \frac{\bar{\nu}\sqrt{T}\sqrt{\Phi_0}}{R\sqrt{\pi B}} \quad , \tag{2.2}$$

with $\Phi_0$ is the unit flux, $R$ is the transverse size of the vortex bundle, $B$ is the value of the applied magnetic field. The essential property can be comprehended by considering the problem of random walk. Let $l_0$ be the length of each individual step, for a walk with totally N steps, if the walk were coherently in the same direction, the total length of N steps would be $L = N\, l_0$ ; however, these steps are not coherently in the same direction, they are at random. With the aim of receiving the length after N steps it must be divided the above expression by the square root of N,

$$L = \frac{N\, l_0}{\sqrt{N}} = \sqrt{N}\, l_0 \quad , \tag{2.3}$$

which is the desired answer for the random walk with N steps.

*2.2. Evaluation the root-mean-square of the angle between the random collective pinning force and positive y-direction*

Let us consider the case for p-type superconductors with current in the positive x-direction and the magnetic field in the positive z-direction, if we assume that the angle between the random collective pinning force of a vortex line and the positive y-direction measured in counterclockwise sense is $\theta$. This temperature and field dependent $\theta$ can be obtained as follows: Since $\theta$ is small, we can approximately write $\theta \cong \frac{|\vec{f}_{el}|}{|\vec{f}_L|}$, where $|\vec{f}_{el}|$ and $|\vec{f}_L|$ are the magnitudes of the elastic force and the Lorentz force of the vortex line. Owing to the thermal fluctuations, taking into account the fact that $C_{11} >> C_{66}$, the displacement vector $\vec{S}_f(\vec{r})$ of the vortex in the bundle as well as its corresponding elastic force $\vec{f}_{el}(\vec{r})$ is proportional to $\sqrt{\frac{k_B}{C_{66}}}$ or $\frac{1}{\sqrt{B}}\sqrt{\frac{T}{T_c - T}}$ [4, 6-9]. The temperature-and-field-dependent $\theta$ can now be expressed as

$$\theta = \bar{\alpha}' \frac{1}{\sqrt{B}} \sqrt{\frac{T}{T_c - T}} \quad , \tag{2.4}$$

where $\bar{\alpha}'$ is a proportional constant. By considering the theorem of random walk, the direction of the random collective pinning force for the bundle is

$$\alpha(T, B) = \sqrt{N}\, \theta = \bar{\alpha} \sqrt{\frac{T}{T_c - T}} \quad , \tag{2.5}$$

with $\bar{\alpha} = \bar{\alpha}' R \sqrt{\pi / \Phi_0}$, where $R$ is the radius of the vortex bundle and $\Phi_0$ is the unit flux. The root-mean-square of the angle between the random collective pinning force and the positive y-direction in counterclockwise sense for the vortex bundle can be obtained as,

$$\sqrt{<\phi^2>_R} = [\int_{-\frac{\pi}{2}}^{\frac{\pi}{2}} \phi^2 \exp(\frac{-\phi^2}{\alpha^2(T, B)})\, d\phi]^{\frac{1}{2}} \quad . \tag{2.6}$$

Keeping in mind the fact that $\alpha(T, B)$ is usually very small in our theory, we obtain,

$$\sqrt{<\phi^2>_R} \cong \alpha(T, B) = \bar{\alpha} \sqrt{\frac{T}{T_c - T}} \quad . \tag{2.7}$$

*2.3. Calculation of the Hall as well as longitudinal resistivity*

To proceed let us calculate the energy barrier of the vortex bundle generated by the Magnus force, the collective pinning force, and the randomly distributed strong pinning sites inside the bundle. After some algebra, the energy barrier of the vortex bundle both in the positive and negative x-direction as well as y-direction are obtained respectively as,

$$U - \overline{V}R(JB\frac{|v_{by}|}{v_T} + < F_{p_x} >_R) \quad , \tag{2.8}$$

$$U + \overline{V}R(JB\frac{|v_{by}|}{v_T} + < F_{p_x} >_R) \quad , \tag{2.9}$$

$$U - \overline{V}R(-JB + JB\frac{v_{bx}}{v_T} + < F_{p_y} >_R) \, , \tag{2.10}$$

$$U + \overline{V}R(-JB + JB\frac{v_{bx}}{v_T} + < F_{p_y} >_R) \quad , \tag{2.11}$$

in which the potential barrier $U$ is generated by the randomly distributed strong pinning sites inside the bundle, $\vec{J} = J\,\vec{e}_x$ is the transport current, $\vec{v}_b$ ($\vec{v}_T$) is the velocity of the vortex- bundle (supercurrent), $< \vec{F}_p >_R$ stands for the random average of the random collective pinning force per unit volume, $\vec{B} = B\vec{e}_z$ is the magnetic field, $\overline{V}$ is the volume of the vortex bundle, $R$ represents the transverse size of the vortex bundle, and the range of $U$ is assumed to be of the order of $R$. The self-consistent equations for the velocity of the thermally activated motion for the vortex bundles over the directional-dependent energy barrier are therefore obtained in components as,

$$v_{bx} = v_C\, R\{\exp[\frac{-1}{k_B T}(U - \overline{V}R(JB\frac{|v_{by}|}{v_T} + < F_{p_x} >_R))]$$

$$-\exp[\frac{-1}{k_B T}(U+\bar{V}R(JB\frac{|v_{by}|}{v_T}+<F_{px}>_R))]\}, \tag{2.12}$$

$$v_{by}=v_C\,R\{\exp[\frac{-1}{k_B T}(U-\bar{V}R(-JB+JB\frac{v_{bx}}{v_T}+<F_{p_y}>_R))]$$

$$-\exp[\frac{-1}{k_B T}(U+\bar{V}R(-JB+JB\frac{v_{bx}}{v_T}+<F_{p_y}>_R))]\} \quad , \tag{2.13}$$

$v_C$ presents the coherent oscillation frequency of the vortex bundle, and the distance between two adjacent minimum potentials of the vortex bundle is assumed to be of the order of $R$.

Observing the fact that $\frac{v_{bx}}{v_T}<<1$, Eqs (2.12) and (2.13) can be approximately rewritten as,

$$v_{bx}=v_C\,R\exp(\frac{-U}{k_B T})\{\exp[\frac{+\bar{V}R}{k_B T}(\frac{JB|v_{by}|}{v_T}+<F_{p_x}>_R)]-\exp[\frac{-\bar{V}R}{k_B T}(\frac{JB|v_{by}|}{v_T}+<F_{p_x}>_R)]\}, \tag{2.14}$$

$$v_{by}=v_C\,R\exp(\frac{-U}{k_B T})\{\exp[\frac{-\bar{V}R}{k_B T}(JB-<F_{p_y}>_R)]-\exp[\frac{\bar{V}R}{k_B T}(JB-<F_{p_y}>_R)]\} \quad , \tag{2.15}$$

together with $\vec{E}=-\vec{v}_b\times\vec{B}$, $\rho_{xx}=\frac{E_x}{J}$, and $\rho_{xy}=\frac{E_y}{J}$, finally $\rho_{xx}$ and $\rho_{xy}$ can be obtained as,

$$\rho_{xx}=\frac{\bar{v}\sqrt{BT\Phi_0}}{J\sqrt{\pi}}\exp(\frac{-U}{k_B T})\{\exp[\frac{\bar{V}R}{k_B T}(JB-(\frac{\beta^C(T,B)}{\bar{V}})^{\frac{1}{2}})]-\exp[\frac{-\bar{V}R}{k_B T}(JB-(\frac{\beta^C(T,B)}{\bar{V}})^{\frac{1}{2}})]\} \quad , \tag{2.16}$$

$$\rho_{xy}=\frac{-\bar{v}\sqrt{BT\Phi_0}}{J\sqrt{\pi}}\exp(\frac{-U}{k_B T})\{\exp[\frac{\bar{V}R}{k_B T}((\frac{\beta^C(T,B)}{\bar{V}})^{\frac{1}{2}}\bar{\alpha}\sqrt{\frac{T}{T_c-T}}-JB\frac{|v_{by}|}{v_T})]$$

$$-\exp[\frac{-\bar{V}R}{k_B T}((\frac{\beta^C(T,B)}{\bar{V}})^{\frac{1}{2}}\bar{\alpha}\sqrt{\frac{T}{T_c-T}}-JB\frac{|v_{by}|}{v_T})]\} \quad , \tag{2.17}$$

$$|v_{by}|=J\rho_{xx}/B \quad , \tag{2.18}$$

where $(\frac{\beta^C(T,B)}{\bar{V}})^{\frac{1}{2}}$ is the magnitude of the random collective pinning force per unit volume [6].

It is well understood that when temperature (magnetic field) below $T_P$ $(B_P)$, the quasiorder-disorder phase transition temperature (magnetic field), the vortex lines form large vortex bundles [6], both Hall and longitudinal resistivities approach to zero quickly as temperature (magnetic field) decreasing. When temperature (magnetic field) above $T_P$ $(B_P)$, the vortex lines form a disordered amorphous vortex system; however, they are not single-quantized vortex lines, the vortex lines still bounded closed together to form small vortex bundles of dimension $R \approx 10^{-8} m$. By pondering the fact that the arguments in the exponential functions inside the curly bracket of Eqs (2.16) and (2.17) are very small when the Lorentz force is closed to the random collective pinning force, Eqs (2.16) and (2.17) can be rewritten as,

$$\rho_{xx} = \frac{\bar{v}\sqrt{\Phi_0}\sqrt{B}}{J\sqrt{\pi}\sqrt{T}} \exp(\frac{-U}{k_B T})\,(\frac{2\bar{V}R}{k_B})\,[JB - (\frac{\beta^C(T,B)}{\bar{V}})^{\frac{1}{2}}] \quad , \qquad (2.19)$$

$$\rho_{xy} = \frac{-\bar{v}\sqrt{\Phi_0}\sqrt{B}}{J\sqrt{\pi}\sqrt{T}} \exp(\frac{-U}{k_B T})\,(\frac{2\bar{V}R}{k_B})\,[(\frac{\beta^C(T,B)}{\bar{V}})^{\frac{1}{2}}\bar{\alpha}\sqrt{\frac{T}{T_c - T}} - JB\frac{|v_{by}|}{v_T}], \qquad (2.20)$$

with $|v_{by}| = J\rho_{xx}/B$. Interesting to stress that, from the above equations, taking into account the fact that $JB > (\frac{\beta^C(T,B)}{\bar{V}})^{\frac{1}{2}}$, for constant applied magnetic field (temperature), the random collective pinning force per unit volume $(\frac{\beta^C(T,B)}{\bar{V}})^{\frac{1}{2}}$ increases with decreasing (increasing) temperature (applied magnetic field). As temperature (applied magnetic field) decreases, the term $JB - (\frac{\beta^C(T,B)}{\bar{V}})^{\frac{1}{2}}$ in Eq (2.19) decreases monotonically; while the factor $(\frac{\beta^C(T,B)}{\bar{V}})^{\frac{1}{2}}\bar{\alpha}\sqrt{\frac{T}{T_c - T}}$ in Eq (2.20) increases, passing the value of $JB\frac{|v_{by}|}{v_T}$, and reaches a maximum, then decreases again. Therefore, $\rho_{xx}$ decreases monotonically and $\rho_{xy}$ decreases crossing over from positive to negative value, reaching a minimum, then increases again. These truly reflect the nature of anomalous Hall effect.

We would like to emphasize that, from the above straightforward calculations, the viscous drag force plays no role in our theory. Our theory is different from the flux-flow theory; in the flux flow theory, the direction of vortex motion with respect to the transport current, the values of $\rho_{xy}$ and $\rho_{xx}$ depended heavily on the balance of the Magnus force and the viscous drag force.

It will become clear in subsequent sections that the AHE includes two regions: the thermally activated motion of small vortex bundles and the thermally activated motion of large vortex bundles; however, the flux-flow region does not belong to it. Also worthily mentioning that in the flux-flow region the pinning centers serve only as the short range scattering centers of order of the coherent length $\xi$ for the vortex core. Besides, these pinning centers are randomly distributed over the sample due to quenched disorder, hence the effects of pinning during the vortex motion tends to be averaged out. Therefore, the over all effects of pinning are insignificant in the flux-flow region.

## 3. The anomalous Hall effect for type-II superconducting bulk materials

In this section we first calculate the longitudinal and Hall resistivities for type-II superconducting bulk materials, in this case, Eqs (2.19) and (2.20) become,

$$\rho_{xx} = \frac{\bar{v}\sqrt{\Phi_0}\sqrt{B}}{J\sqrt{\pi}\sqrt{T}} \exp(\frac{-U}{k_B T})\ (\frac{2\pi R^3 L}{k_B})\,[JB - (\frac{\beta^C(T,B)}{\bar{V}})^{\frac{1}{2}}] \quad , \tag{3.1}$$

$$\rho_{xy} = \frac{-\bar{v}\sqrt{\Phi_0}\sqrt{B}}{J\sqrt{\pi}\sqrt{T}} \exp(\frac{-U}{k_B T})[\frac{2\pi R^3 L}{k_B}][(\frac{\beta^C(T,B)}{\bar{V}})^{\frac{1}{2}}\bar{\alpha}\sqrt{\frac{T}{T_c - T}} - JB\frac{|v_{by}|}{v_T}] \quad , \tag{3.2}$$

with $|v_{by}| = J\rho_{xx} / B$. As we have indicated before, the above equations give rise to the phenomenon of AHE, namely, as the applied magnetic field (temperature) decreases, the value of $\rho_{xx}$ decreases monotonically, and $\rho_{xy}$ decreases crossing over to negative value and reaches its minimum value then increases again. We will discuss the cases both for constant temperature and constant applied magnetic field in the following subsections separately.

*3.1. The Hall and longitudinal resistivities for constant temperature*

In this subsection we study the case when the temperature of the system is at a constant value $T$. As we have declared before, when applied magnetic field decreases, the value of $\rho_{xy}$ decreases, crossing over to negative value and reaches its minimum, then increases again. It is interesting to compare with the experiments for this case. When the temperature is kept at a constant $T = 91K$, from Eqs (3.1) and (3.2) together with Eq (2.1), the Hall and the longitudinal resistivities as functions of applied magnetic field in Tesla are given as follows:

$\rho_{xy}(B=4) = 2.57\times10^{-9}\Omega - m$, $\rho_{xy}(3.03) = 9.23\times10^{-11}\ \Omega - m$, $\rho_{xy}(2) = -3.71\times10^{-9}\Omega - m$,

$\rho_{xy}(1) = -1.23\times10^{-9}\Omega - m$; $\rho_{xx}(4) = 3.02\times10^{-6}\Omega - m$, $\rho_{xx}(3.03) = 2.04\times10^{-6}\ \Omega - m$,

$\rho_{xx}(2) = 6.97\times10^{-7}\Omega - m$, $\rho_{xx}(1) = 3.52\times10^{-7}\Omega - m$ .

In obtaining the above results, the following numbers have been employed approximately,

$R = 2\times10^{-8} m$, $L = 10^{-6} m$, $J = 10^{6}\dfrac{A}{m^2}$, $T_C = 91.7K$, $v_T = 10^{3} m/\sec$, $\overline{\alpha} = 3.76\times10^{-5} T^{\frac{-1}{2}}$,

$\overline{\nu} = 10^{11}\sec^{-1}$, $\exp(\dfrac{-U}{k_B T}) = 3.87\times10^{-3}$, $(\dfrac{\beta^C(B=4)}{\overline{V}})^{\frac{1}{2}} = 3.6\times10^{6} N/m^3$,

$(\dfrac{\beta^C(3.03)}{\overline{V}})^{\frac{1}{2}} = 2.72\times10^{6} N/m^3$, $(\dfrac{\beta^C(2)}{\overline{V}})^{\frac{1}{2}} = 1.87\times10^{6} N/m^3$, $(\dfrac{\beta^C(1)}{\overline{V}})^{\frac{1}{2}} = 9.07\times10^{5} N/m^3$.

From the above results we demonstrate that the value of $\rho_{xx}$ decreases monotonically and $\rho_{xy}$ indeed decreases crossing over from positive to negative and reaches its minimum value, then increases again with decreasing the applied field. If the applied field decreases beyond 1 Tesla, from our previous study, the quasiorder-disorder phase transition occurs [6]; this is the region of the thermally activated motion for the large bundles, both the Hall and longitudinal resistivties approach to zero quickly as the applied field decreasing. These results are in good consistent with experimental data on $YBa_2Cu_3O_{7-\delta}$ high-Tc bulk materials [20]. The regime crosses over to the usual flux-flow regime, when the field increases beyond 4 Tesla.

*3.2. The Hall and longitudinal resistivities for constant applied magnetic field*

Considering the case when the applied magnetic field is kept at a constant value $B = 2.24T$, the Hall and the longitudinal resistivities as functions of temperature are given numerically as,

$\rho_{xy}(T = 91.3K) = 6.3\times10^{-12}\Omega - m$, $\rho_{xy}(91) = -5.527\times10^{-12}\ \Omega - m$, $\rho_{xy}(89) = -6.726\times10^{-9}\Omega - m$,

$\rho_{xy}(87) = -5.67\times10^{-9}\Omega - m$; $\rho_{xx}(91.3) = 3.27\times10^{-6}\Omega - m$, $\rho_{xx}(91) = 2.067\times10^{-6}\ \Omega - m$,

$\rho_{xx}(89) = 2.53\times10^{-7}\Omega - m$, $\rho_{xx}(87) = 1.31\times10^{-7}\Omega - m$.

In arriving the above results, the following data are approximately used,

$R = 2\times10^{-8} m$ , $L = 10^{-6} m$ , $J = 10^{6}\frac{A}{m^{2}}$, $T_C = 92K$ , $v_T = 10^{3} m/\sec$ , $\overline{\alpha} = 1.2\times10^{-5} T^{\frac{-1}{2}}$ ,

$\overline{\nu} = 10^{11}\sec^{-1}$ , $\exp(\frac{-U}{k_B T}) = 4.06\times10^{-3}$ , $(\frac{\beta^C(T = 91.3)}{\overline{V}})^{\frac{1}{2}} = 1.69\times10^{6} N/m^{3}$ ,

$(\frac{\beta^C(91)}{\overline{V}})^{\frac{1}{2}} = 1.893\times10^{6} N/m^{3}$ , $(\frac{\beta^C(89)}{\overline{V}})^{\frac{1}{2}} = 2.198\times10^{6} N/m^{3}$ , $(\frac{\beta^C(87)}{\overline{V}})^{\frac{1}{2}} = 2.22\times10^{6} N/m^{3}$ .

Shown from the above results, the value of $\rho_{xx}$ decreases monotonically, and $\rho_{xy}$ is genuinely crossing over from positive to negative value and reaches its minimum, then increases again with decreasing the temperature. As the temperature is less than $87K$ , according to our previous study, the quasiorder-disorder phase transition occurs [6], it crosses over to the region of thermally activated motion for large vortex bundles, both the values of $\rho_{xy}$ and $\rho_{xx}$ move toward zero quickly as the temperature decreasing. These results are in good agreement with the experimental data on $YBa_2Cu_3O_{7-\delta}$ high-Tc bulk materials [20]. The regime crosses over to the usual flux-flow regime, when temperature is greater than $91.3K$ .

## 4. The anomalous Hall effect for type-II superconducting films

For type-II superconducting films, as we have detailed discussed in Sec 2.3, from Eqs (2.19) and (2.20), $\rho_{xx}$ and $\rho_{xy}$ now become,

$$\rho_{xx} = \frac{\overline{\nu}\sqrt{\Phi_0}\sqrt{B}}{J\sqrt{\pi}\sqrt{T}}\exp(\frac{-U}{k_B T})\,(\frac{2\pi R^3 d}{k_B})\,[JB - (\frac{\beta^C(T,B)}{\overline{V}})^{\frac{1}{2}}] \quad , \tag{4.1}$$

$$\rho_{xy} = \frac{-\overline{\nu}\sqrt{\Phi_0}\sqrt{B}}{J\sqrt{\pi}\sqrt{T}}\exp(\frac{-U}{k_B T})[\frac{2\pi R^3 d}{k_B}][(\frac{\beta^C(T,B)}{\overline{V}})^{\frac{1}{2}}\overline{\alpha}\sqrt{\frac{T}{T_c - T}} - JB\frac{|v_{by}|}{v_T}], \tag{4.2}$$

with $|v_{by}| = J\rho_{xx}/B$ . As discussed in Section 3, the above equations sincerely signify the phenomenon of AHE for both the constant applied magnetic field as well as constant temperature. We will examine the cases in the following subsections.

*4.1. The Hall and longitudinal resistivities for constant temperature*

Studying the case when the temperature is kept at a constant value $T = 4.5K$ , the values of $\rho_{xy}$ and $\rho_{xx}$ as functions of applied magnetic field in Tesla are given numerically as follows:

$\rho_{xy}(B = 7.25) = 1.311\times10^{-12}\Omega - m$ , $\rho_{xy}(7) = -1.5\times10^{-11}\ \Omega - m$ , $\rho_{xy}(6.25) = -1.404\times10^{-10}\Omega - m$ ,

$\rho_{xy}(6) = -7.66\times10^{-11}\Omega - m$ , $\rho_{xx}(7.25) = 8.15\times10^{-7}\ \Omega - m$ , $\rho_{xx}(7) = 7.667\times10^{-7}\Omega - m$ ,

$\rho_{xx}(6.25) = 6.05\times10^{-7}\Omega - m$ , $\rho_{xx}(6) = 5.978\times10^{-7}\Omega - m$ .

Where the following data have been employed approximately, $R = 2\times10^{-8}m$ , $d = 5\times10^{-8}m$ ,

$J = 5\times10^{5}\dfrac{A}{m^{2}}$ , $T_C = 7.5K$ , $v_T = 10^{2}m/\sec$ , $\overline{\alpha} = 1.168\times10^{-3}T^{\frac{-1}{2}}$ , $\overline{\nu} = 10^{11}\sec^{-1}$ ,

$\exp(\dfrac{-U}{k_B T}) = 3.085\times10^{-4}$ , $(\dfrac{\beta^{C}(B = 7.25)}{\overline{V}})^{\frac{1}{2}} = 1.399\times10^{6}N/m^{3}$ , $(\dfrac{\beta^{C}(7)}{\overline{V}})^{\frac{1}{2}} = 1.369\times10^{6}N/m^{3}$ ,

$(\dfrac{\beta^{C}(6.25)}{\overline{V}})^{\frac{1}{2}} = 1.364\times10^{6}N/m^{3}$ , $(\dfrac{\beta^{C}(6)}{\overline{V}})^{\frac{1}{2}} = 1.205\times10^{6}N/m^{3}$ .

Analyzing above results, we see that the value of $\rho_{xx}$ decreases monotonically and the $\rho_{xy}$ decreases crossing over from positive to negative value and reaches its minimum, then increases again as the applied magnetic field decreasing, When the applied magnetic field decreasing below 6 Tesla, the quasiorder-disorder phase transition takes place [6], belonging to the region of thermally activated motion of large vortex bundles, both the values of $\rho_{xy}$ and $\rho_{xx}$ approach to zero rapidly as the applied magnetic field decreasing. The results are in good agreement with the experimental data on $Mo_3Si$ low-Tc thin films [21]. When the applied magnetic field greater than $7.25$ Tesla, it belongs to the usual flux-flow regime.

*4.2. The Hall and longitudinal resistivities for constant applied magnetic field*

Next focusing the situation when the applied field $B$ is kept at a constant value $B = 1T$, the values of $\rho_{xy}$ and $\rho_{xx}$ as functions of temperature become,

$\rho_{xy}(T = 92.6K) = 1.87\times 10^{-9}\Omega - m$, $\rho_{xy}(92) = 7.072\times 10^{-12}\ \Omega - m$, $\rho_{xy}(91) = -2.81\times 10^{-9}\Omega - m$, $\rho_{xy}(90.75) = -2.73\times 10^{-9}\Omega - m$; $\rho_{xx}(92.6) = 4.89\times 10^{-7}\Omega - m$, $\rho_{xx}(92) = 3.301\times 10^{-7}\ \Omega - m$, $\rho_{xx}(91) = 8.371\times 10^{-8}\Omega - m$, $\rho_{xx}(90.75) = 7.86\times 10^{-8}\Omega - m$ .

In deriving the above results, the following data are used approximately, $R = 2\times 10^{-8} m$, $d = 5\times 10^{-8} m$, $J = 10^6 \frac{A}{m^2}$, $T_C = 94K$, $v_T = 10^2 m/\sec$, $\bar{\alpha} = 7.352\times 10^{-4} T^{\frac{-1}{2}}$, $\bar{\nu} = 10^{11}\sec^{-1}$,

$\exp(\frac{-U}{k_B T}) = 2.01\times 10^{-2}$, $(\frac{\beta^C(T = 92.6)}{\bar{V}})^{\frac{1}{2}} = 4.99\times 10^5 N/m^3$, $(\frac{\beta^C(92)}{\bar{V}})^{\frac{1}{2}} = 6.63\times 10^5 N/m^3$

$(\frac{\beta^C(91)}{\bar{V}})^{\frac{1}{2}} = 9.15\times 10^5 N/m^3$, $(\frac{\beta^C(90.75)}{\bar{V}})^{\frac{1}{2}} = 9.203\times 10^5 N/m^3$.

Recognizing the above results, the value of $\rho_{xx}$ decreases monotonically and $\rho_{xy}$ decreases crossing over from positive to negative value and reaches its minimum, then increases again as the temperature decreasing; the quasiorder-disorder phase transition occurs [6] when the temperature decreasing below $87K$, it then crosses over to the region of thermally activated motion of the large vortex bundles, both $\rho_{xy}$ and $\rho_{xx}$ come close to zero quickly as the temperature decreasing. Our results do exhibit the expected anomalous nature and are in consistence with experimental data on $YBa_2Cu_3O_{7-\delta}$ high-Tc thin films [22]. When the temperature greater than $91.3K$, it crosses over to the usual flux-flow region.

## 5. The reentry phenomenon for anomalous Hall effect

In this section we investigate the situation that if the random collective pinning force $(\frac{\beta^C(T,B)}{\bar{V}})^{\frac{1}{2}}$ is not too large, the value of $\rho_{xx}$ decreases monotonically and $\rho_{xy}$ decreases crossing over from positive to negative value at $T_a$, the onset temperature for sign reversal, reaching a minimum, then increases crossing over from negative back to positive value at $T_R$, the reentry temperature, reaching a maximum value then decreases again with decreasing temperature. This is the fascinating reentry phenomenon. The condition for occurring the reentry is that if the random collective pinning force is not too large, such that when $T_P < T_R$, where $T_P$ is the temperature for the quasiorder-disorder first-order phase transition for the sample, then the reentry phenomenon could happen; if $(\frac{\beta^C(T,B)}{\bar{V}})^{\frac{1}{2}}$ is large enough, such that when $T_P > T_R$, then the reentry effect could not be observed. This is the reason why $YBa_2Cu_3O_{7-\delta}$ high-Tc materials do not have reentry effect at high applied magnetic field [23]. To make numerical estimations for the above case, let us consider the applied field is kept at $B = 2T$, the Hall and the longitudinal resistivities as functions of temperature are given as,

$\rho_{xy}(T = 102K) = 1.724\times10^{-11}\Omega - m$, $\rho_{xy}(100) = 6.22\times10^{-12}$ $\Omega - m$, $\rho_{xy}(96) = -7.73\times10^{-13}\Omega - m$,

$\rho_{xy}(85) = -8.201\times10^{-14}$ $\Omega - m$, $\rho_{xy}(76) = 3.003\times10^{-13}\Omega - m$, $\rho_{xy}(74) = 2.82\times10^{-13}\Omega - m$;

$\rho_{xx}(102) = 2.195\times10^{-8}\Omega - m$, $\rho_{xx}(100) = 1.108\times10^{-8}\ \Omega - m$, $\rho_{xx}(96) = 3.37\times10^{-9}\Omega - m$,

$\rho_{xx}(85) = 2.529\times10^{-9}\Omega - m$, $\rho_{xx}(76) = 2.317\times10^{-9}\Omega - m$, $\rho_{xx}(74) = 2.21\times10^{-9}\Omega - m$.

In obtaining the above results, the following approximate numbers have been used,

$R = 2\times10^{-8}m$, $L = 10^{-6}m$, $J = 5\times10^{6}\dfrac{A}{m^2}$, $T_C = 104K$, $v_T = 10^2 m/\sec$, $\overline{\alpha} = 3.278\times10^{-5}T^{\frac{-1}{2}}$,

$\overline{\nu} = 10^{11}\sec^{-1}$, $\exp(\dfrac{-U}{k_B T}) = 1.47\times10^{-5}$, $(\dfrac{\beta^C(T=102)}{\overline{V}})^{\frac{1}{2}} = 4.3\times10^{6}N/m^3$, $(\dfrac{\beta^C(100)}{\overline{V}})^{\frac{1}{2}} = 7.15\times10^{6}N/m^3$

$(\dfrac{\beta^C(96)}{\overline{V}})^{\frac{1}{2}} = 9.15\times10^{6}N/m^3$ $(\dfrac{\beta^C(85)}{\overline{V}})^{\frac{1}{2}} = 9.4\times10^{6}N/m^3$, $(\dfrac{\beta^C(76)}{\overline{V}})^{\frac{1}{2}} = 9.48\times10^{6}N/m^3$,

$(\dfrac{\beta^C(74)}{\overline{V}})^{\frac{1}{2}} = 9.51\times10^{6}N/m^3$.

Our present results shown that the value of $\rho_{xx}$ decreases monotonically and the $\rho_{xy}$ decreases and crossing over from positive to negative value and reaches its minimum, then increases crossing over from negative back to positive value, researching a small local maximum then decreases again as the temperature decreasing. For temperature decreasing below $72K$, the quasiorder-disorder first-order phase transition happens [6]. The system turns into the region of thermally activated motion of the large vortex bundles, both the values of $\rho_{xy}$ and $\rho_{xx}$ approach to zero promptly as the temperature decreasing. Our results reflect the fascinating reentry phenomenon and is in consistence with the experimental data on $Tl_2Ba_2Cu_2O_8$ high-Tc bulk materials [24]. While for the temperature higher than $102K$, the regime crosses over to the usual flux-flow regime.

## 6. Discussion

We would like to point out first of all that we have developed a novel theory for the thermally activated motion of vortex-bundle over the directional-dependent energy barrier formed by potential due to the Magnus force and the random collective pinning force and the potential $U$ generated by the randomly distributed strong pinning sites inside the bundle. We have successfully applied this theory and shown that the AHE is induced by the competition between the Magnus force and the random collective pinning force.

Secondly, our theory for AHE is indeed a microscopic one. Starting from the microscopic theory of BCS [26] by Gor'kov [27], one can get the Ginzgurg-Landau theory [28]. By the work of Abrikosov [1], the flux lines in mixed state type-II superconductors form a long-range order of the FLL. However, in the presence of quenched disorder, the long-range order of the FLL is destroyed, only short-range order prevails [5-6]. Based on the thermally activated motion of the vortex bundles over the directional-dependent potential barrier together with the Maxwell's equation and the effect of random collective pinning [6-9], we arrive at the theory of AHE. Therefore, our theory is a microscopic one.

Thirdly, we would like to emphasize that the quasiorder-disorder first-order phase transition or the peak effect [6] occurs inside the region of AHE. The phenomenon of AHE includes two interesting regions: the thermally activated motion of small vortex bundles, and the thermally activated motion of large vortex bundles; however, the flux flow region does not belong to it. It is worthwhile to stress that we only consider the systems that are in steady state, any time-dependent behavior of the system will not be discussed.

Fourthly, our theory is a very general theory, it can be applied to conventional and high-Tc superconductors. Although their mechanisms, the structure of vortex lattice, and even the method of pairing are entirely different, these do not affect the results of our theory.

Finally, the dimensional fluctuations have no effect on the essential structure of our theory. This is why we discuss the 2D and 3D anomalous Hall effect at the same time.

## 7. Conclusion

We have developed a theory for the thermally activated motion over the directional-dependent energy barrier of vortex bundles for the mixed state type-II superconductors, and applied this theory to investigate the AHE for conventional and high-Tc superconductors, bulk materials and thin films. Our calculations demonstrate that the Hall anomaly is indeed induced by the competition between the Magnus force and the random collective pinning force. The AHE for constant applied magnetic field, constant temperature, the reentry phenomenon are also discussed. All the results are in good agreement with the experiments.

**Acknowledgements**

W Y Chen would like to thank Professor S Feng for useful and constructive discussions.